\def\papertitle{A Frequency-Domain Artificial Reverberator Plug-In}
\def\paperauthorA{Jonas Roth}
\def\paperauthorB{Nishanth Kumar}
\def\paperauthorC{Silvan Krebs}
\def\paperauthorD{David Wieland}
\def\paperauthorE{Christoph Studer}

\documentclass[twoside,a4paper]{article}
\usepackage{etoolbox}

\usepackage{dafx26demo}

\usepackage{amsmath,amssymb,amsfonts,amsthm}
\usepackage{siunitx}
\usepackage{euscript}
\usepackage[T1]{fontenc}
\usepackage[utf8]{inputenc}
\usepackage{ifpdf}
\usepackage[english]{babel}
\usepackage{caption}
\usepackage{subcaption}
\usepackage{color}
\usepackage{booktabs}

\input glyphtounicode
\ninept

\usepackage{amssymb}
\usepackage{amsfonts}
\usepackage{mathrsfs}
\usepackage{xspace}
\usepackage{bm}
\usepackage{upgreek}

\newcommand{\safemath}[2]{\newcommand{#1}{\ensuremath{#2}\xspace}}

\safemath{\bma}{\mathbf{a}}
\safemath{\bmb}{\mathbf{b}}
\safemath{\bmc}{\mathbf{c}}
\safemath{\bmd}{\mathbf{d}}
\safemath{\bme}{\mathbf{e}}
\safemath{\bmf}{\mathbf{f}}
\safemath{\bmg}{\mathbf{g}}
\safemath{\bmh}{\mathbf{h}}
\safemath{\bmi}{\mathbf{i}}
\safemath{\bmj}{\mathbf{j}}
\safemath{\bmk}{\mathbf{k}}
\safemath{\bml}{\mathbf{l}}
\safemath{\bmm}{\mathbf{m}}
\safemath{\bmn}{\mathbf{n}}
\safemath{\bmo}{\mathbf{o}}
\safemath{\bmp}{\mathbf{p}}
\safemath{\bmq}{\mathbf{q}}
\safemath{\bmr}{\mathbf{r}}
\safemath{\bms}{\mathbf{s}}
\safemath{\bmt}{\mathbf{t}}
\safemath{\bmu}{\mathbf{u}}
\safemath{\bmv}{\mathbf{v}}
\safemath{\bmw}{\mathbf{w}}
\safemath{\bmx}{\mathbf{x}}
\safemath{\bmy}{\mathbf{y}}
\safemath{\bmz}{\mathbf{z}}
\safemath{\bmzero}{\mathbf{0}}
\safemath{\bmone}{\mathbf{1}}

\bmdefine{\biad}{a}
\bmdefine{\bibd}{b}
\bmdefine{\bicd}{c}
\bmdefine{\bidd}{d}
\bmdefine{\bied}{e}
\bmdefine{\bifd}{f}
\bmdefine{\bigd}{g}
\bmdefine{\bihd}{h}
\bmdefine{\biid}{i}
\bmdefine{\bijd}{j}
\bmdefine{\bikd}{k}
\bmdefine{\bild}{l}
\bmdefine{\bimd}{m}
\bmdefine{\bind}{n}
\bmdefine{\biod}{o}
\bmdefine{\bipd}{p}
\bmdefine{\biqd}{q}
\bmdefine{\bird}{r}
\bmdefine{\bisd}{s}
\bmdefine{\bitd}{t}
\bmdefine{\biud}{u}
\bmdefine{\bivd}{v}
\bmdefine{\biwd}{w}
\bmdefine{\bixd}{x}
\bmdefine{\biyd}{y}
\bmdefine{\bizd}{z}

\bmdefine{\bixid}{\xi}
\bmdefine{\bilambdad}{\lambda}
\bmdefine{\bimud}{\mu}
\bmdefine{\bithetad}{\theta}
\bmdefine{\biphid}{\phi}
\bmdefine{\bideltad}{\delta}

\safemath{\bmia}{\biad}
\safemath{\bmib}{\bibd}
\safemath{\bmic}{\bicd}
\safemath{\bmid}{\bidd}
\safemath{\bmie}{\bied}
\safemath{\bmif}{\bifd}
\safemath{\bmig}{\bigd}
\safemath{\bmih}{\bihd}
\safemath{\bmii}{\biid}
\safemath{\bmij}{\bijd}
\safemath{\bmik}{\bikd}
\safemath{\bmil}{\bild}
\safemath{\bmim}{\bimd}
\safemath{\bmin}{\bind}
\safemath{\bmio}{\biod}
\safemath{\bmip}{\bipd}
\safemath{\bmiq}{\biqd}
\safemath{\bmir}{\bird}
\safemath{\bmis}{\bisd}
\safemath{\bmit}{\bitd}
\safemath{\bmiu}{\biud}
\safemath{\bmiv}{\bivd}
\safemath{\bmiw}{\biwd}
\safemath{\bmix}{\bixd}
\safemath{\bmiy}{\biyd}
\safemath{\bmiz}{\bizd}

\safemath{\bmxi}{\bixid}
\safemath{\bmlambda}{\bilambdad}
\safemath{\bmmu}{\bimud}
\safemath{\bmtheta}{\bithetad}
\safemath{\bmphi}{\biphid}
\safemath{\bmdelta}{\bideltad}

\safemath{\bA}{\mathbf{A}}
\safemath{\bB}{\mathbf{B}}
\safemath{\bC}{\mathbf{C}}
\safemath{\bD}{\mathbf{D}}
\safemath{\bE}{\mathbf{E}}
\safemath{\bF}{\mathbf{F}}
\safemath{\bG}{\mathbf{G}}
\safemath{\bH}{\mathbf{H}}
\safemath{\bI}{\mathbf{I}}
\safemath{\bJ}{\mathbf{J}}
\safemath{\bK}{\mathbf{K}}
\safemath{\bL}{\mathbf{L}}
\safemath{\bM}{\mathbf{M}}
\safemath{\bN}{\mathbf{N}}
\safemath{\bO}{\mathbf{O}}
\safemath{\bP}{\mathbf{P}}
\safemath{\bQ}{\mathbf{Q}}
\safemath{\bR}{\mathbf{R}}
\safemath{\bS}{\mathbf{S}}
\safemath{\bT}{\mathbf{T}}
\safemath{\bU}{\mathbf{U}}
\safemath{\bV}{\mathbf{V}}
\safemath{\bW}{\mathbf{W}}
\safemath{\bX}{\mathbf{X}}
\safemath{\bY}{\mathbf{Y}}
\safemath{\bZ}{\mathbf{Z}}

\safemath{\bZero}{\mathbf{0}}
\safemath{\bOne}{\mathbf{1}}
\safemath{\bDelta}{\mathbf{\Delta}}
\safemath{\bLambda}{\mathbf{\UpLambda}}
\safemath{\bPhi}{\mathbf{\Upphi}}
\safemath{\bSigma}{\mathbf{\Upsigma}}
\safemath{\bOmega}{\mathbf{\Upomega}}
\safemath{\bTheta}{\mathbf{\Uptheta}}

\bmdefine{\biAd}{A}
\bmdefine{\biBd}{B}
\bmdefine{\biCd}{C}
\bmdefine{\biDd}{D}
\bmdefine{\biEd}{E}
\bmdefine{\biFd}{F}
\bmdefine{\biGd}{G}
\bmdefine{\biHd}{H}
\bmdefine{\biId}{I}
\bmdefine{\biJd}{J}
\bmdefine{\biKd}{K}
\bmdefine{\biLd}{L}
\bmdefine{\biMd}{M}
\bmdefine{\biOd}{N}
\bmdefine{\biPd}{O}
\bmdefine{\biQd}{P}
\bmdefine{\biRd}{R}
\bmdefine{\biSd}{S}
\bmdefine{\biTd}{T}
\bmdefine{\biUd}{U}
\bmdefine{\biVd}{V}
\bmdefine{\biWd}{W}
\bmdefine{\biXd}{X}
\bmdefine{\biYd}{Y}
\bmdefine{\biZd}{Z}

\bmdefine{\biDelta}{\Delta}
\bmdefine{\biLambda}{\Lambda}
\bmdefine{\biPhi}{\Phi}
\bmdefine{\biSigma}{\Sigma}
\bmdefine{\biOmega}{\Omega}
\bmdefine{\biTheta}{\Theta}

\safemath{\bimA}{\biAd}
\safemath{\bimB}{\biBd}
\safemath{\bimC}{\biCd}
\safemath{\bimD}{\biDd}
\safemath{\bimE}{\biEd}
\safemath{\bimF}{\biFd}
\safemath{\bimG}{\biGd}
\safemath{\bimH}{\biHd}
\safemath{\bimI}{\biId}
\safemath{\bimJ}{\biJd}
\safemath{\bimK}{\biKd}
\safemath{\bimL}{\biLd}
\safemath{\bimM}{\biMd}
\safemath{\bimN}{\biNd}
\safemath{\bimO}{\biOd}
\safemath{\bimP}{\biPd}
\safemath{\bimQ}{\biQd}
\safemath{\bimR}{\biRd}
\safemath{\bimS}{\biSd}
\safemath{\bimT}{\biTd}
\safemath{\bimU}{\biUd}
\safemath{\bimV}{\biVd}
\safemath{\bimW}{\biWd}
\safemath{\bimX}{\biXd}
\safemath{\bimY}{\biYd}
\safemath{\bimZ}{\biZd}

\safemath{\bimDelta}{\biDelta}
\safemath{\bimLambda}{\biLambda}
\safemath{\bimPhi}{\biPhi}
\safemath{\bimSigma}{\biSigma}
\safemath{\bimOmega}{\biOmega}
\safemath{\bimTheta}{\biTheta}

\safemath{\setA}{\mathcal{A}}
\safemath{\setB}{\mathcal{B}}
\safemath{\setC}{\mathcal{C}}
\safemath{\setD}{\mathcal{D}}
\safemath{\setE}{\mathcal{E}}
\safemath{\setF}{\mathcal{F}}
\safemath{\setG}{\mathcal{G}}
\safemath{\setH}{\mathcal{H}}
\safemath{\setI}{\mathcal{I}}
\safemath{\setJ}{\mathcal{J}}
\safemath{\setK}{\mathcal{K}}
\safemath{\setL}{\mathcal{L}}
\safemath{\setM}{\mathcal{M}}
\safemath{\setN}{\mathcal{N}}
\safemath{\setO}{\mathcal{O}}
\safemath{\setP}{\mathcal{P}}
\safemath{\setQ}{\mathcal{Q}}
\safemath{\setR}{\mathcal{R}}
\safemath{\setS}{\mathcal{S}}
\safemath{\setT}{\mathcal{T}}
\safemath{\setU}{\mathcal{U}}
\safemath{\setV}{\mathcal{V}}
\safemath{\setW}{\mathcal{W}}
\safemath{\setX}{\mathcal{X}}
\safemath{\setY}{\mathcal{Y}}
\safemath{\setZ}{\mathcal{Z}}
\safemath{\emptySet}{\varnothing}

\safemath{\colA}{\mathscr{A}}
\safemath{\colB}{\mathscr{B}}
\safemath{\colC}{\mathscr{C}}
\safemath{\colD}{\mathscr{D}}
\safemath{\colE}{\mathscr{E}}
\safemath{\colF}{\mathscr{F}}
\safemath{\colG}{\mathscr{G}}
\safemath{\colH}{\mathscr{H}}
\safemath{\colI}{\mathscr{I}}
\safemath{\colJ}{\mathscr{J}}
\safemath{\colK}{\mathscr{K}}
\safemath{\colL}{\mathscr{L}}
\safemath{\colM}{\mathscr{M}}
\safemath{\colN}{\mathscr{N}}
\safemath{\colO}{\mathscr{O}}
\safemath{\colP}{\mathscr{P}}
\safemath{\colQ}{\mathscr{Q}}
\safemath{\colR}{\mathscr{R}}
\safemath{\colS}{\mathscr{S}}
\safemath{\colT}{\mathscr{T}}
\safemath{\colU}{\mathscr{U}}
\safemath{\colV}{\mathscr{V}}
\safemath{\colW}{\mathscr{W}}
\safemath{\colX}{\mathscr{X}}
\safemath{\colY}{\mathscr{Y}}
\safemath{\colZ}{\mathscr{Z}}

\safemath{\opA}{\mathbb{A}}
\safemath{\opB}{\mathbb{B}}
\safemath{\opC}{\mathbb{C}}
\safemath{\opD}{\mathbb{D}}
\safemath{\opE}{\mathbb{E}}
\safemath{\opF}{\mathbb{F}}
\safemath{\opG}{\mathbb{G}}
\safemath{\opH}{\mathbb{H}}
\safemath{\opI}{\mathbb{I}}
\safemath{\opJ}{\mathbb{J}}
\safemath{\opK}{\mathbb{K}}
\safemath{\opL}{\mathbb{L}}
\safemath{\opM}{\mathbb{M}}
\safemath{\opN}{\mathbb{N}}
\safemath{\opO}{\mathbb{O}}
\safemath{\opP}{\mathbb{P}}
\safemath{\opQ}{\mathbb{Q}}
\safemath{\opR}{\mathbb{R}}
\safemath{\opS}{\mathbb{S}}
\safemath{\opT}{\mathbb{T}}
\safemath{\opU}{\mathbb{U}}
\safemath{\opV}{\mathbb{V}}
\safemath{\opW}{\mathbb{W}}
\safemath{\opX}{\mathbb{X}}
\safemath{\opY}{\mathbb{Y}}
\safemath{\opZ}{\mathbb{Z}}
\safemath{\opZero}{\mathbb{O}}
\safemath{\identityop}{\opI}

\safemath{\veca}{\bma}
\safemath{\vecb}{\bmb}
\safemath{\vecc}{\bmc}
\safemath{\vecd}{\bmd}
\safemath{\vece}{\bme}
\safemath{\vecf}{\bmf}
\safemath{\vecg}{\bmg}
\safemath{\vech}{\bmh}
\safemath{\veci}{\bmi}
\safemath{\vecj}{\bmj}
\safemath{\veck}{\bmk}
\safemath{\vecl}{\bml}
\safemath{\vecm}{\bmm}
\safemath{\vecn}{\bmn}
\safemath{\veco}{\bmo}
\safemath{\vecp}{\bmp}
\safemath{\vecq}{\bmq}
\safemath{\vecr}{\bmr}
\safemath{\vecs}{\bms}
\safemath{\vect}{\bmt}
\safemath{\vecu}{\bmu}
\safemath{\vecv}{\bmv}
\safemath{\vecw}{\bmw}
\safemath{\vecx}{\bmx}
\safemath{\vecy}{\bmy}
\safemath{\vecz}{\bmz}

\safemath{\veczero}{\bmzero}
\safemath{\vecone}{\bmone}
\safemath{\vecxi}{\bmxi}
\safemath{\veclambda}{\bmlambda}
\safemath{\vecmu}{\bmmu}
\safemath{\vectheta}{\bmtheta}
\safemath{\vecphi}{\bmphi}
\safemath{\vecdelta}{\bmdelta}

\safemath{\matA}{\bA}
\safemath{\matB}{\bB}
\safemath{\matC}{\bC}
\safemath{\matD}{\bD}
\safemath{\matE}{\bE}
\safemath{\matF}{\bF}
\safemath{\matG}{\bG}
\safemath{\matH}{\bH}
\safemath{\matI}{\bI}
\safemath{\matJ}{\bJ}
\safemath{\matK}{\bK}
\safemath{\matL}{\bL}
\safemath{\matM}{\bM}
\safemath{\matN}{\bN}
\safemath{\matO}{\bO}
\safemath{\matP}{\bP}
\safemath{\matQ}{\bQ}
\safemath{\matR}{\bR}
\safemath{\matS}{\bS}
\safemath{\matT}{\bT}
\safemath{\matU}{\bU}
\safemath{\matV}{\bV}
\safemath{\matW}{\bW}
\safemath{\matX}{\bX}
\safemath{\matY}{\bY}
\safemath{\matZ}{\bZ}
\safemath{\matzero}{\bmzero}

\safemath{\matDelta}{\bDelta}
\safemath{\matLambda}{\bLambda}
\safemath{\matPhi}{\bPhi}
\safemath{\matSigma}{\bSigma}
\safemath{\matOmega}{\bOmega}
\safemath{\matTheta}{\bTheta}

\safemath{\matidentity}{\matI}
\safemath{\matone}{\matO}

\safemath{\rnda}{A}
\safemath{\rndb}{B}
\safemath{\rndc}{C}
\safemath{\rndd}{D}
\safemath{\rnde}{E}
\safemath{\rndf}{F}
\safemath{\rndg}{G}
\safemath{\rndh}{H}
\safemath{\rndi}{I}
\safemath{\rndj}{J}
\safemath{\rndk}{K}
\safemath{\rndl}{L}
\safemath{\rndm}{M}
\safemath{\rndn}{N}
\safemath{\rndo}{O}
\safemath{\rndp}{P}
\safemath{\rndq}{Q}
\safemath{\rndr}{R}
\safemath{\rnds}{S}
\safemath{\rndt}{T}
\safemath{\rndu}{U}
\safemath{\rndv}{V}
\safemath{\rndw}{W}
\safemath{\rndx}{X}
\safemath{\rndy}{Y}
\safemath{\rndz}{Z}

\safemath{\rveca}{\bimA}
\safemath{\rvecb}{\bimB}
\safemath{\rvecc}{\bimC}
\safemath{\rvecd}{\bimD}
\safemath{\rvece}{\bimE}
\safemath{\rvecf}{\bimF}
\safemath{\rvecg}{\bimG}
\safemath{\rvech}{\bimH}
\safemath{\rveci}{\bimI}
\safemath{\rvecj}{\bimJ}
\safemath{\rveck}{\bimK}
\safemath{\rvecl}{\bimL}
\safemath{\rvecm}{\bimM}
\safemath{\rvecn}{\bimN}
\safemath{\rveco}{\bomO}
\safemath{\rvecp}{\bimP}
\safemath{\rvecq}{\bimQ}
\safemath{\rvecr}{\bimR}
\safemath{\rvecs}{\bimS}
\safemath{\rvect}{\bimT}
\safemath{\rvecu}{\bimU}
\safemath{\rvecv}{\bimV}
\safemath{\rvecw}{\bimW}
\safemath{\rvecx}{\bimX}
\safemath{\rvecy}{\bimY}
\safemath{\rvecz}{\bimZ}

\safemath{\rvecxi}{\bmxi}
\safemath{\rveclambda}{\bmlambda}
\safemath{\rvecmu}{\bmmu}
\safemath{\rvectheta}{\bmtheta}
\safemath{\rvecphi}{\bmphi}

\safemath{\rmatA}{\bimA}
\safemath{\rmatB}{\bimB}
\safemath{\rmatC}{\bimC}
\safemath{\rmatD}{\bimD}
\safemath{\rmatE}{\bimE}
\safemath{\rmatF}{\bimF}
\safemath{\rmatG}{\bimG}
\safemath{\rmatH}{\bimH}
\safemath{\rmatI}{\bimI}
\safemath{\rmatJ}{\bimJ}
\safemath{\rmatK}{\bimK}
\safemath{\rmatL}{\bimL}
\safemath{\rmatM}{\bimM}
\safemath{\rmatN}{\bimN}
\safemath{\rmatO}{\bimO}
\safemath{\rmatP}{\bimP}
\safemath{\rmatQ}{\bimQ}
\safemath{\rmatR}{\bimR}
\safemath{\rmatS}{\bimS}
\safemath{\rmatT}{\bimT}
\safemath{\rmatU}{\bimU}
\safemath{\rmatV}{\bimV}
\safemath{\rmatW}{\bimW}
\safemath{\rmatX}{\bimX}
\safemath{\rmatY}{\bimY}
\safemath{\rmatZ}{\bimZ}

\safemath{\rmatDelta}{\bimDelta}
\safemath{\rmatLambda}{\bimLambda}
\safemath{\rmatPhi}{\bimPhi}
\safemath{\rmatSigma}{\bimSigma}
\safemath{\rmatOmega}{\bimOmega}
\safemath{\rmatTheta}{\bimTheta}

\usepackage{amssymb}
\usepackage{amsfonts}
\usepackage{mathrsfs}
\usepackage{xspace}
\usepackage{bm}
\usepackage{fancyref}
\usepackage{textcomp}

\usepackage{multirow}
\usepackage{stmaryrd}

\newenvironment{textbmatrix}{	\setlength{\arraycolsep}{2.5pt}%
								\big[\begin{matrix}}{\end{matrix}\big]%
								\raisebox{0.08ex}{\vphantom{M}}}

\def\be{\begin{equation}}
\def\ee{\end{equation}}
\def\een{\nonumber \end{equation}}
\def\mat{\begin{bmatrix}}
\def\emat{\end{bmatrix}}
\def\btm{\begin{textbmatrix}}
\def\etm{\end{textbmatrix}}

\def\ba#1\ea{\begin{align}#1\end{align}}
\def\bas#1\eas{\begin{align*}#1\end{align*}}
\def\bs#1\es{\begin{split}#1\end{split}}
\def\bg#1\eg{\begin{gather}#1\end{gather}}
\def\bml#1\eml{\begin{multline}#1\end{multline}}
\def\bi#1\ei{\begin{itemize}#1\end{itemize}}

\safemath{\dirac}{\delta}					
\safemath{\krond}{\dirac}					

\safemath{\upto}{\uparrow}
\safemath{\downto}{\downarrow}
\safemath{\iu}{j}							
\safemath{\ev}{\lambda}						
\safemath{\hilseqspace}{l^{2}}				
\newcommand{\banachfunspace}[1]{\setL^{#1}}	
\safemath{\hilfunspace}{\banachfunspace{2}}	

\safemath{\SNR}{\textit{SNR}} 				
\safemath{\PAR}{\textit{PAR}} 				
\safemath{\No}{N_0}							
\safemath{\Es}{E_s}							
\safemath{\Eb}{E_b}							
\safemath{\EbNo}{\frac{\Eb}{\No}}
\safemath{\EsNo}{\frac{\Es}{\No}}

\DeclareMathOperator{\CHop}{\ensuremath{\opH}} 
\safemath{\tvir}{\rndh_{\CHop}}				
\safemath{\tvtf}{\rndl_{\CHop}}				
\safemath{\spf}{\rnds_{\CHop}}				
\safemath{\bff}{H_{\CHop}}					

\safemath{\ircf}{r_{h}}						
\safemath{\tftvcf}{r_{s}}					
\safemath{\tfcf}{r_{l}}						
\safemath{\bfcf}{r_{H}}						

\safemath{\tcorr}{c_h}						
\safemath{\scf}{c_{s}}						
\safemath{\tfcorr}{c_{l}}					
\safemath{\fcorr}{c_{H}}						

\safemath{\mi}{I}							
\safemath{\capacity}{C}						

\safemath{\normal}{\mathcal{N}}			
\safemath{\jpg}{\mathcal{CN}}			
\safemath{\mchain}{\leftrightarrow}		

\safemath{\dB}{\,\mathrm{dB}}
\safemath{\dBm}{\,\mathrm{dBm}}
\safemath{\Hz}{\,\mathrm{Hz}}
\safemath{\kHz}{\,\mathrm{kHz}}
\safemath{\MHz}{\,\mathrm{MHz}}
\safemath{\GHz}{\,\mathrm{GHz}}
\safemath{\s}{\,\mathrm{s}}
\safemath{\ms}{\,\mathrm{ms}}
\safemath{\mus}{\,\mathrm{\text{\textmu}s}}
\safemath{\ns}{\,\mathrm{ns}}
\safemath{\ps}{\,\mathrm{ps}}
\safemath{\meter}{\,\mathrm{m}}
\safemath{\mm}{\,\mathrm{mm}}
\safemath{\cm}{\,\mathrm{cm}}
\safemath{\m}{\,\mathrm{m}}
\safemath{\W}{\,\mathrm{W}}
\safemath{\mW}{\, \mathrm{mW}}
\safemath{\J}{\,\mathrm{J}}
\safemath{\K}{\,\mathrm{K}}
\safemath{\bit}{\,\mathrm{bit}}
\safemath{\nat}{\,\mathrm{nat}}

\safemath{\define}{\triangleq}			

\safemath{\equivalent}{\sim}
\safemath{\distas}{\sim}					
\safemath{\sdiff}{\Delta}				

\safemath{\reals}{\mathbb{R}}
\safemath{\positivereals}{\reals_{+}}
\safemath{\integers}{\mathbb{Z}}
\safemath{\posint}{\integers_{+}}
\safemath{\naturals}{\mathbb{N}}
\safemath{\posnaturals}{\naturals_{+}}
\safemath{\complexset}{\mathbb{C}}
\safemath{\rationals}{\mathbb{Q}}

\newcommand*{\fancyrefapplabelprefix}{app}		
\newcommand*{\fancyrefthmlabelprefix}{thm}		
\newcommand*{\fancyreflemlabelprefix}{lem}		
\newcommand*{\fancyrefcorlabelprefix}{cor}		
\newcommand*{\fancyrefdeflabelprefix}{def}		
\newcommand*{\fancyrefproplabelprefix}{prop}		
\newcommand*{\fancyrefexmpllabelprefix}{exmpl}
\newcommand*{\fancyrefalglabelprefix}{alg}		
\newcommand*{\fancyreftbllabelprefix}{tbl}		

\frefformat{vario}{\fancyrefseclabelprefix}{Section~#1}
\frefformat{vario}{\fancyrefthmlabelprefix}{Theorem.~#1}
\frefformat{vario}{\fancyreftbllabelprefix}{Table~#1}
\frefformat{vario}{\fancyreflemlabelprefix}{Lemma~#1}
\frefformat{vario}{\fancyrefcorlabelprefix}{Corollary~#1}
\frefformat{vario}{\fancyrefdeflabelprefix}{Definition~#1}
\frefformat{vario}{\fancyreffiglabelprefix}{Figure~#1}
\frefformat{vario}{\fancyrefapplabelprefix}{Appendix~#1}
\frefformat{vario}{\fancyrefeqlabelprefix}{(#1)}
\frefformat{vario}{\fancyrefproplabelprefix}{Proposition~#1}
\frefformat{vario}{\fancyrefexmpllabelprefix}{Example~#1}
\frefformat{vario}{\fancyrefalglabelprefix}{Algorithm~#1}

 \newtheorem*{remark*}{Remark}

\safemath{\dictab}{[\,\dicta\,\,\dictb\,]}

\safemath{\ysig}{\bmy}
\safemath{\ysighat}{\hat{\ysig}}
\safemath{\ysigdim}{M}
\safemath{\xsig}{\bmx}
\safemath{\xsigdim}{N}
\safemath{\nx}{n_x}
\safemath{\zsig}{\bmz}
\safemath{\zsigdim}{\ysigdim}
\safemath{\rsig}{\bmr}
\safemath{\Adict}{\bA}
\safemath{\Adicttilde}{\widetilde{\Adict}}
\safemath{\Adictdim}{\outputdim\times\xsigdim}
\safemath{\avec}{\bma}
\safemath{\avectilde}{\tilde{\avec}}
\safemath{\Bdict}{\bB}
\safemath{\Bdicttilde}{\widetilde{\Bdict}}
\safemath{\Cdict}{\bC}
\safemath{\cvec}{\bmc}
\safemath{\Ddict}{\bD}
\safemath{\Ddictdim}{\ysigdim\times\xsigdim}
\safemath{\dvec}{\bmd}
\safemath{\Ddicttilde}{\widetilde{\bD}}
\safemath{\Bonb}{\bB}
\safemath{\bvec}{\bmb}
\safemath{\Bonbdim}{\ysigdim\times\ysigdim}
\safemath{\noise}{\bmn}
\safemath{\noisedim}{\ysigim}
\safemath{\err}{\bme}
\safemath{\errdim}{\ysigdim}
\safemath{\errset}{\setE}
\safemath{\nerr}{n_e}
\safemath{\delop}{\bP_\errset}
\safemath{\delopc}{\bP_{{\errset}^c}}

\safemath{\cplxi}{\imath}
\safemath{\cplxj}{\jmath}

\safemath{\dict}{\matD}
\safemath{\inputdim}{N}		
\safemath{\outputdim}{M}		
\safemath{\sparsity}{S}	
\safemath{\inputdimA}{{N_a}}	
\safemath{\inputdimB}{{N_b}}	
\safemath{\elemA}{{n_a}}	
\safemath{\elemB}{{n_b}}	
\safemath{\resA}{\matR_a}	
\safemath{\resB}{\matR_b}	
\safemath{\subD}{\matS} 
\safemath{\subA}{\matS_a} 
\safemath{\subB}{\matS_b} 
\safemath{\dicta}{\matA} 	
\safemath{\dictb}{\matB} 	
\safemath{\hollowS}{H}
\safemath{\hollowA}{H_a}
\safemath{\hollowB}{H_b}
\safemath{\cross}{Z}
\safemath{\coh}{\mu_d}			
\safemath{\coha}{\mu_a}			
\safemath{\cohb}{\mu_b}			
\safemath{\mubs}{\nu}	
\safemath{\cohm}{\mu_m} 
\safemath{\dictset}{\setD}	
\safemath{\dictsetp}{\dictset(\coh,\coha,\cohb)}	
\safemath{\dictsetgen}{\dictset_\text{gen}}
\safemath{\dictsetgenp}{\dictsetgen(\coh)}
\safemath{\dictsetonb}{\dictset_\text{onb}}
\safemath{\dictsetonbp}{\dictsetonb(\coh)}

\safemath{\leftside}{U}
\safemath{\rightsideA}{R_a}
\safemath{\rightsideB}{R_b}

\safemath{\indexS}{\setI_S} 

\safemath{\na}{n_a}			
\safemath{\nb}{n_b}			
\safemath{\coeffa}{p_i}	
\safemath{\coeffb}{q_j}	
\safemath{\seta}{\setP}		
\safemath{\setb}{\setQ}     
\safemath{\setw}{\setW}	
\safemath{\setz}{\setZ}	
\safemath{\cola}{\veca}		
\safemath{\colb}{\vecb}		
\safemath{\cold}{\vecd}		
\safemath{\inputvec}{\vecx} 	
\safemath{\error}{\vece}	
\safemath{\noiseout}{\vecz} 	
\safemath{\inputvecel}{x}
\safemath{\inputveca}{\vecx_a}
\safemath{\inputvecb}{\vecx_b}
\safemath{\outputvec}{\vecy}	
\safemath{\lambdamin}{\lambda_{\mathrm{min}}}

\safemath{\elltwo}{\ell_2}
\safemath{\ellone}{\ell_1}
\safemath{\ellzero}{\ell_0}
\safemath{\ellinf}{\ell_\infty}
\safemath{\ellinftilde}{\ell_{\widetilde\infty}}
\safemath{\licard}{Z(\coh,\coha,\cohb)}
\safemath{\xsol}{\hat{x}}
\safemath{\xbord}{x_b}		
\safemath{\xstat}{x_s}		
\safemath{\xstatLone}{\tilde{x}_s}
\safemath{\order}{\mathcal{O}} 
\safemath{\scales}{\Theta} 
\safemath{\ones}{\mathbf{1}} 
\safemath{\zeroes}{\mathbf{0}} 
\safemath{\thlone}{\kappa(\coh,\cohb)} 
\safemath{\constoneA}{\delta} 
\safemath{\constoneB}{\epsilon} 
\safemath{\nlarge}{L}				   
\safemath{\sumlarge}{S_\nlarge}
\safemath{\maxlarger}{P_\nlarge}	   
\safemath{\Pzero}{\textrm{P0}}	
\safemath{\Pone}{\textrm{P1}}
\safemath{\vecfir}{\vecw}			 
\safemath{\vecsec}{\vecz}
\safemath{\elvecfir}{w}              
\safemath{\elvecsec}{z}				 
\safemath{\nlargefir}{n}
\safemath{\normout}{\gamma}
\safemath{\auxfun}{h}
\safemath{\supp}{\textrm{supp}}

\safemath{\indexa}{\ell}
\safemath{\indexb}{r}
\safemath{\indexc}{i}
\safemath{\indexd}{j}

\safemath{\project}{P}
\usepackage{cuted} 
\usepackage{microtype}

\newcommand{\fdverb}{FDverb\xspace}
\newcommand{\param}[1]{\texttt{#1}\xspace}

\newcommand{\cur}[0]{[\ell]}
\newcommand{\prev}[0]{[\ell\!-\!1]}

\newcommand{\envinbin}[1][k]{\varepsilon_{#1}}

\newcommand{\envbin}[1][k]{\bar{\varepsilon}_{#1}}

\newcommand{\attcoef}[1][k]{\alpha_{#1}}
\newcommand{\relcoef}[1][k]{\beta_{#1}}

\newcommand{\relcoefarith}[1][k]{\beta'_{#1}}
\newcommand{\ofststat}[0]{\sigma}
\newcommand{\ofstdyn}[0]{\delta}

\newcommand{\her}[0]{h_\mathrm{ER}}
\newcommand{\predelay}[0]{\ensuremath{T_\textrm{pre}}\xspace}
\newcommand{\taildelay}[0]{\ensuremath{T_\textrm{tail}}\xspace}
\newcommand{\taildelaymin}[0]{\ensuremath{T_\textrm{tail, min}}\xspace}
\newcommand{\Tproc}[0]{\ensuremath{T_\textrm{proc}}\xspace}
\newcommand{\Talgo}[0]{\ensuremath{T_\textrm{algo}}\xspace}
\newcommand{\erdecaytime}[0]{T_\textrm{60,ER}}
\newcommand{\Tsixty}[0]{\ensuremath{T_\textrm{60}}\xspace}

\newcommand{\fs}{f_\mathrm{s}}

\newcounter{numauth}
\newcounter{listcnt}
\newcommand\authcnt[1]{\ifdefined#1 \stepcounter{numauth} \fi}

\newcommand\addauth[1]{
\ifdefined#1
\stepcounter{listcnt}
\ifnum \value{listcnt}<\value{numauth}
\appto\authorslist{, #1}
\else
\appto\authorslist{~and~#1}
\fi
\fi}
\authcnt{\paperauthorB}
\authcnt{\paperauthorC}
\authcnt{\paperauthorD}
\authcnt{\paperauthorE}
\authcnt{\paperauthorF}
\authcnt{\paperauthorG}
\authcnt{\paperauthorH}
\authcnt{\paperauthorI}
\authcnt{\paperauthorJ}
\def\authorslist{\paperauthorA}
\addauth{\paperauthorB}
\addauth{\paperauthorC}
\addauth{\paperauthorD}
\addauth{\paperauthorE}
\addauth{\paperauthorF}
\addauth{\paperauthorG}
\addauth{\paperauthorH}
\addauth{\paperauthorI}
\addauth{\paperauthorJ}

\usepackage{times}

\newif\ifpdf
\ifx\pdfoutput\relax
\else
   \ifcase\pdfoutput
      \pdffalse
   \else
      \pdftrue
   \fi
\fi

\ifpdf 
  \usepackage[pdftex,
    pdftitle={\papertitle},
    pdfauthor={\authorslist},
    pdfsubject={Proceedings of the 29th International Conference on Digital Audio Effects (DAFx26)},
    colorlinks=false, 
    bookmarksnumbered, 
    pdfstartview=XYZ 
  ]{hyperref}
  \usepackage[pdftex]{graphicx}
\else 
  \usepackage[dvips]{epsfig,graphicx}
  \usepackage[dvips,
    pdftitle={\papertitle},
    pdfauthor={\authorslist},
    pdfsubject={Proceedings of the 29th International Conference on Digital Audio Effects (DAFx26)},
    colorlinks=false, 
    bookmarksnumbered, 
    pdfstartview=XYZ 
  ]{hyperref}
\fi
\usepackage[hypcap=true]{caption}
\title{\papertitle}

\affiliation
{\paperauthorA, \paperauthorB, \paperauthorC, \paperauthorD, and \paperauthorE}
{Department of Information Technology and Electrical Engineering, ETH Zurich, Switzerland\\
{\tt
\href{mailto:joroth@ethz.ch}{joroth@ethz.ch} |
\href{mailto:studer@ethz.ch}{studer@ethz.ch}}
}

\usepackage{standalone}
\usepackage{tikz}
\usepackage{pgfplots}
\pgfplotsset{compat=1.18}

\begin{document}
\ifpdf 
  \DeclareGraphicsExtensions{.png,.jpg,.pdf}
\else  
  \DeclareGraphicsExtensions{.eps}
\fi


\maketitle

\sloppy

\begin{abstract}
We present \fdverb, a frequency-domain artificial reverberator, based on the idea of a vocoder with a noise carrier signal.
Using a short-time Fourier transform (STFT) for analysis and synthesis, \fdverb generates late reverberation by weighting spectral noise components with envelopes.
We extend \fdverb with early reflections, nonlinear decay, and pitch shifting.
These extensions enable creative sound-design applications.
We provide \fdverb as an open-source DAW plug-in, using the JUCE framework.
\end{abstract}

\section{Introduction}
\label{sec:intro}


Artificial reverberation is a fundamental tool in audio production. Although traditionally used to simulate the acoustic properties of physical spaces, artificial reverberators are increasingly valued as creative effects that produce reverberation with no physical counterpart.
The acoustic response of a physical room is typically divided into two parts: (i) early reflections (ERs), where individual reflections are still distinguishable, and (ii) late reverberation tail, where the reflection density is high and the response takes on a stochastic character.
%
As observed in \cite{moorer1979about}, some impulse responses of concert halls sound remarkably similar to white noise with an exponential decay of amplitude.
This observation motivated synthetic reverberation methods such as \cite{vickers2006frequency, vilkamo2010sparse}, which model late reverberation as noise with frequency-dependent decay.
More specifically, \cite{vickers2006frequency} describes a frequency-domain reverberator based on recursive spectral magnitude decay, derived from the phase vocoder.
Related,~\cite{vilkamo2010sparse} adopts bandwise exponentially decaying noise as an ideal reference and proposes a computationally efficient frequency-domain reverberator that is perceptually equivalent to the ideal reference.
There also exists a number of so-called ``spectral processing'' DAW plug-ins, such as \cite{waproduction_venom}, that seem to implement similar techniques, but technical details are not available.

\subsection{Contributions}

Inspired by the idea of a vocoder with a noise carrier, we derive a late reverberation generator, further referred to as \emph{tail reverberator}.
The tail reverberator operates in the frequency domain using a short-time Fourier transform (STFT), and is conceptually related to the spectral magnitude decay reverberator of~\cite{vickers2006frequency}.
In addition, we implement nonlinear decay and pitch shifting for the tail reverberator, which enables creative sound-design applications.
Finally, we combine the tail reverberator with an early-reflection unit; together, they form our proposed hybrid reverberation algorithm, which we call \emph{\fdverb}.
An early version of \fdverb was demonstrated on \emph{graetli}, a microcontroller-based audio DSP platform~\cite{dafx2025graetli}, but the underlying algorithm was not described.
We provide \fdverb as an open-source DAW plug-in together with a Python reference implementation, both available in the \fdverb Git repository.\footnote{\url{https://github.com/IIP-Group/FDverb/}}

\section{The FDverb Algorithm}
\label{sec:algorithm}

\begin{figure}[tp]
\centering
\includegraphics[width=8cm]{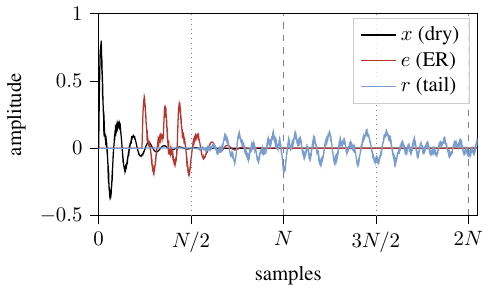}
\caption{\label{fig:dry_er_tail}Illustration of signal components of \fdverb applied to a drum signal: (black) input $x$, (red) early reflections $e$, and (blue) reverb tail $r$. The dashed and dotted vertical lines mark the boundaries of the overlapping blocks processed by the tail reverberator (here, we used a block size of $N = 8192$).}
\end{figure}

\begin{figure*}[tp]
\centering
\includegraphics[scale=1.0]{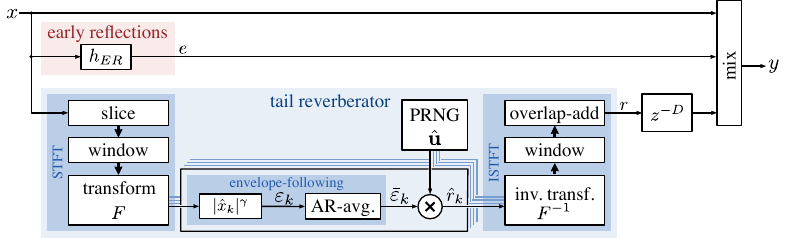}
\caption{\label{fig:blockdiagram}Block diagram for \fdverb.}
\end{figure*}

We now explain the hybrid \fdverb algorithm that contains an ER unit and a tail reverberator, as depicted in \fref{fig:blockdiagram}.
\fdverb receives a discrete-time \emph{dry} input signal $x[n],~n \in \mathbb{Z}$ (we will use the shorthand $x$).
The ER unit generates the ER signal $e$ by convolving~$x$ with a sparse impulse response, as described in \fref{sec:early_reflections}.
The tail reverberator generates the tail signal $r$ by shaping the spectrum of a noise source to match the spectrum of the input signal $x$, as described in \fref{sec:reverb_tail}.
The \fdverb output signal $y$ is a weighted sum of $x$, $e$, and a delayed version of $r$; \fref{fig:dry_er_tail} illustrates these three components for a drum signal.
In the following discussion, we display the parameters of the \fdverb user interface in a monospaced font, e.g., \param{parameter-name}.

\subsection{Early Reflections}
\label{sec:early_reflections}

We obtain the ER signal $e$ by convolving the input $x$ with a sparse early-reflection impulse response (ER-IR) $\her$.
We implement the convolution in the time domain to avoid the block latency inherent to frequency-domain processing, and constrain the impulse response $\her$ to be sparse and of finite length, in order to limit both the computational cost and the memory usage.

More specifically, we synthesize $\her$ based on the following parameters:
(i) $E$~(\param{echoes}), the number of early reflections to be generated;
(ii) \predelay~(\param{pre-delay}), the delay of the first early reflection;
(iii) \taildelay~(\param{tail-delay}), the maximum delay across all early reflections;
and
(iv) $\erdecaytime$, the ER decay time\footnote{In the \fdverb plug-in, $\erdecaytime$ is not exposed in the user interface, but is derived from the nominal \param{decay} time parameter instead.}.

First, we draw the tap delays randomly from a quadratic distribution over the interval $[\predelay,\taildelay]$, except the first tap, which is fixed at \predelay.
Based on the tap delays, we compute the tap gains to decay exponentially, with decay time $\erdecaytime$.
We implement the random draw of the tap delays using a pseudo-random number generator (PRNG) and use \predelay as the seed for the PRNG, so that the user gets repeatable results for the impulse response $\her$.

\subsection{Reverb-Tail Algorithm}
\label{sec:reverb_tail}

At a high level, the \fdverb tail reverberator implements the following steps:
\begin{enumerate}
\item Apply STFT to the tail-reverberator input signal $x$.
\item For each spectral bin $k$, extract the magnitude envelope $\envbin$.
\item Generate the noise carrier spectrum $\hat{\bmu}$.
\item For each spectral bin, modulate the amplitude of the noise with the envelope, to get the reverb-tail spectrum bins $\hat{r}_k$.
\item Apply inverse STFT (ISTFT) to $\hat{\bmr}$, which leads to the tail-reverberator output signal $r$.
\end{enumerate}

In the following, we detail a mono implementation of \fdverb; later we will extend \fdverb to a stereo implementation.

\subsubsection{Forward Short-Time Fourier Transform}
\label{sec:forward_transform}

Consider the signal $x$, which enters the tail algorithm on the left side in \fref{fig:blockdiagram}.
We realize the STFT by slicing the input signal into overlapping blocks of $N$ samples, which we denote as $\bmx\cur \in \mathbb{R}^N$, where $\ell$ is used to index the blocks over time.
We use an overlap of~$\frac{N}{2}$ between the blocks and, for simplicity, constrain $N$ to powers of two.
To reduce blocking artifacts, we apply a sine window to each block.
Next, we apply a spectral transform $F(\bmx\cur) = \hat\bmx\cur \in \mathbb{C}^K$ from $N$ time-domain samples to $K$ frequency bins.
For our explanation, we consider the case of a unitary \emph{discrete Fourier transform} (DFT), where we only consider the non-negative frequency bins, so $K = \frac{N}{2}+1$.\footnote{When implemented using a fast Fourier transform (FFT) algorithm, the above transform is sometimes referred to as a ``real-FFT''~(RFFT).}

\subsubsection{Envelope-Following}

We now perform envelope-following for each bin $\hat{x}_k, k=1,\ldots,K$.
The envelope follower comprises a \emph{detector} followed by an \emph{averager}.
The detector is implemented as follows.
For each block $\ell$, we compute the detector signal $\envinbin\cur \in \mathbb{R}$ for the bin $k$ as
\begin{align}
\envinbin\cur &= \left|\hat{x}_k\cur\right|^\gamma,
\end{align}
where $|\cdot|$ denotes the magnitude and $\gamma > 0$ shapes the amplitude dynamics of the reverb tail.
The default value $\gamma = 1$ corresponds to amplitude detection.
In the user interface, we refer to $\gamma$ as the \param{color} parameter.\footnote{When the color parameter is set to $\gamma \neq 1$, then the detector also influences the spectral balance of the tail signal, since typical input signals have a $1/f$ spectral distribution.}


The averager is implemented as a so-called \emph{attack-release (AR)-averager}: a first-order IIR low-pass filter with two distinct time constants for the attack and release phases of the signal $\envbin\cur$, as described in \cite{zolzer2011dafx,mcnally1984dynamic}.
For each block $\ell$, we compute the AR-averager signal $\envbin \in \mathbb{R}$ for the bin $k$ as
\begin{align}\label{eq:envelope-exp}
\envbin\cur\!&=\!\left\{\begin{array}{@{}l@{}l@{~}l@{}}
\attcoef \,\envinbin\cur & + (1\!-\!\attcoef) \envbin\prev &\text{ if }~\envinbin\cur \geq \envbin\prev\!\!\!\\
\relcoef \,\envinbin\cur & + (1\!-\!\relcoef) \envbin\prev &\text{ if }~\envinbin\cur < \envbin\prev,\!\!\!\!\!
\end{array}\right.
\end{align}
where the coefficients $\attcoef, \relcoef \in (0,1)$ control the attack- and release-time constants of the envelope, respectively.

Through the user interface, the release-time coefficients $\relcoef$ are controlled as follows.
The \param{decay} parameter specifies a nominal decay time constant (given as $\Tsixty$).
We then use the \param{tilt} parameter to derive a frequency-dependent decay time constant $\Tsixty(k)$ for each bin, from which we determine $\relcoef$.\footnote{We set $\relcoef = 1-10^{\frac{-3\,N/2}{\fs\,\Tsixty(k)}}$, where $\Tsixty(k) \in (0,\,2\,\Tsixty]$.}
The attack-time coefficient, on the other hand, is not exposed to the user: we found that varying the attack time does not noticeably affect the character of the reverb tail.
We therefore fix $\attcoef = \attcoef[]$, independent of frequency, to achieve an attack time of $\qty{10}{\ms}$.

\begin{figure}[t]
\centering
\includegraphics[width=\columnwidth]{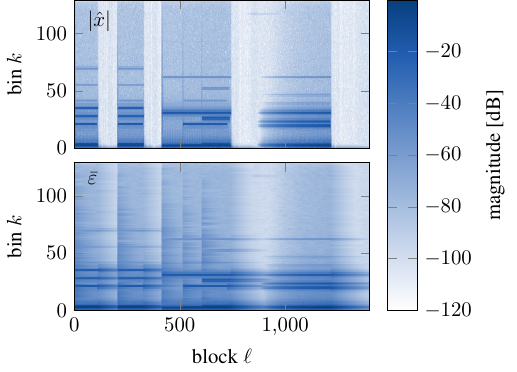}
\caption{\label{fig:env_follow}Time-frequency magnitudes before and after the envelope follower ($N=256$, decay time \qty{750}{\ms}), for an organ-chords input. Top: input magnitude $|\hat{x}|$ (spectrogram); bottom: AR-averager signal $\bar{\varepsilon}$ (``envelope-gram'').}
\vspace{-2mm}
\end{figure}

\fref{fig:env_follow} illustrates the effect of the envelope follower on an input signal containing a sequence of organ chords, whose distinct harmonics are visible as horizontal bands in the spectrogram.
Comparing the input (top) and output (bottom), the AR-averager smooths the signal in time, i.e., along the horizontal direction.

\subsubsection{Spectral Noise Generation and Weighting}

In a conventional vocoder-reverberator setup, one would generate noise in the time domain and apply the same spectral decomposition as for the modulator signal.
For \fdverb, since the DFT of a noise signal will be noise, we generate it directly in the spectral domain.
For each block $\ell$, we generate a noise spectrum $\hat{u}_k \in \mathbb{R},\; k=1,\ldots,K$ as uniformly distributed samples
$\hat{u}_k\cur \overset{\text{i.i.d.}}{\sim} \setU_{[-1,1]}$.


Next, we use the extracted envelopes $\envbin$ from \fref{eq:envelope-exp} to weight the spectral noise bins.
For each block $\ell$, we compute the reverb-tail spectrum as $\hat{r}_k\cur = \hat{u}_k\cur \cdot \envbin\cur$ for $k=1,\ldots,K.$

\vspace{-2mm}
\subsubsection{Inverse Short-Time Fourier Transform}

With the weighted noise $\hat\bmr$, we apply the inverse spectral transform $F^{-1}(\hat\bmr\cur) = \bmr\cur \in \mathbb{R}^N$ from $K$ bins to $N$ samples.
We apply a sine window again to reduce blocking artifacts and then perform overlap-adding of consecutive blocks with an overlap of $\frac{N}{2}$, to obtain the time-domain reverb-tail signal $r$.

\vspace{-2mm}
\subsection{The Impact of Block Size $N$}
\label{sec:block_size}

We discuss two important repercussions of the block size $N$: (i) the effect on the sound character of the reverb tail and (ii) the latency introduced in a real-time implementation.

Together with the sample rate $\fs$, $N$ sets the frequency resolution $\fs/N$ and the block duration $N/\fs$, two reciprocal quantities that trade off against each other.
A large block size (i.e., large $N/\fs$) provides a fine frequency resolution and a tonal tail at the cost of temporal smearing, whereas a short block (small $N/\fs$) provides a coarse frequency resolution and a noise-like tail that follows the input more sharply in time.

In a real-time implementation, block-wise processing delays the tail signal relative to the input, since the input must be buffered and processed before the corresponding tail emerges.
This delay has two contributions: the per-block processing time \Tproc, which is small (a few milliseconds or less), and the algorithmic latency $\Talgo = \frac{N}{2\,\fs}$.
Their sum is the total latency, which imposes a lower bound on the delay between the dry signal and the reverb tail; we denote this bound by $\taildelaymin = \Tproc + \Talgo$, the minimum possible tail-delay.\footnote{For example, $N=8192$ at $\fs=\qty{48}{\kHz}$ results in $\Talgo\approx\qty{85}{\ms}$.}

Without early reflections, this minimum tail-delay would also be the onset of the reverb itself and, hence, the perceived pre-delay; for large $N$, such a long pre-delay audibly detaches the reverb from the dry signal and sounds unnatural.
In \fdverb, we use early reflections to avoid this issue: implemented by time-domain convolution (see \fref{sec:early_reflections}), their pre-delay has no such lower limit, so the reverb can begin with a short pre-delay even when the tail arrives late.
To provide the user with independent control over the pre- and tail-delays, we delay the tail signal so that it aligns with $\taildelay$, the position of the last early reflection.
In the user interface, $\taildelaymin$ then acts as the lower limit of the \param{tail-delay} parameter.


\subsection{Adding Bells and Whistles}
\label{sec:bells_n_whistles}

We now introduce several enhancements that extend the basic \fdverb tail algorithm with additional creative controls.

\subsubsection{Alternative Envelope Modes}

The reverb tail of a physical room typically follows an exponential decay curve (which gave rise to \fref{eq:envelope-exp}, further referred to as \emph{exponential} envelope mode).
Beyond this physically motivated envelope mode, \fdverb offers two additional envelope modes.

In the \emph{arithmetic} envelope mode, we replace \fref{eq:envelope-exp} by
\begin{align}\label{eq:envelope-arith}
\envbin\cur \!&=\!\left\{\begin{array}{@{}l@{}l@{~}l@{}}
\min\{\envbin\prev+\attcoef, \envinbin\cur\}      &\text{ if }~\envinbin\cur\!\geq\!\envbin\prev\!\!\!\!\!\\
\max\{0, \envbin\prev - \relcoefarith, \envinbin\cur\} &\text{ if }~\envinbin\cur\!<\!\envbin\prev\!\!\!\!\!\!
\end{array}\right.
\end{align}
where $\relcoefarith$ are the release-time coefficients for the arithmetic mode.\footnote{We set $\relcoefarith = 4  \frac{N/2}{\fs\,T_1}$, where $T_1$ is the time required for a decrease in amplitude of $1$ and the factor $4$ is an empirical choice.}
The arithmetic envelope mode produces a reverb tail that sounds unnatural and resembles \emph{gated reverb} (sometimes also referred to as \emph{nonlinear decay} reverb).

In the \emph{freeze} mode, we replace \fref{eq:envelope-exp} by
\begin{align}\label{eq:envelope-forever}
\envbin\cur &= \max\{\envinbin\cur,\; \envbin\prev\},
\end{align}
which results in a reverb tail that does not decay, i.e., is sustained infinitely.
This mode is similar to \emph{time-freeze} described in \cite{vickers2006frequency}.

In the \fdverb user interface, the alternative envelope modes are controlled as follows:
The \param{mode} selector chooses between \param{exponential} and \param{arithmetic} envelope modes, while the \param{freeze} button allows one to momentarily enable the freeze envelope mode.
The freeze envelope mode also features an input gate, which is controlled with the \param{gate} toggle.
With the gate \param{open}, the input spectrum is continuously accumulated into the reverb-tail envelopes; with the gate \param{closed}, the tail-reverberator input is muted and the reverb-tail envelopes are held fixed.
The dry signal~$x$ is unaffected in either case.

\subsubsection{Pitch Shift and Pitch Drift}

Taking creative freedom further, we extend \fdverb with two pitch-shifting features for the reverb tail.

With the \param{shift} parameter, we offset the bins of the analysis side and the synthesis side of the tail algorithm. 
In detail, we introduce an input bin offset $\ofststat \in \mathbb{Z}$ and modify the envelope update equations (i.e., \fref{eq:envelope-exp}, \fref{eq:envelope-arith}, \fref{eq:envelope-forever}) to use $\envinbin[k-\ofststat]$ in place of $\envinbin[k]$, treating out-of-bounds elements as zero.
The offset $\ofststat$ applies a static pitch shift to the reverb tail (upward or downward, depending on the sign of $\ofststat$).
However, since this shift is linear in frequency, harmonic intervals are not preserved, resulting in an inharmonic reverb tail.

With the \param{drift} parameter, we offset the state-holding elements of the AR-averager units in the algorithm.
In detail, we introduce a state bin offset $\ofstdyn \in \mathbb{Z}$ and modify the envelope update equations (i.e., \fref{eq:envelope-exp}, \fref{eq:envelope-arith}, \fref{eq:envelope-forever}) to use $\envbin[k-\ofstdyn]\prev$ in place of $\envbin[k]\prev$, treating out-of-bounds elements as zero.
The offset $\ofstdyn$ applies a dynamic pitch shift to the reverb tail, i.e., the pitch of the reverb tail rises or falls over time, depending on the sign of $\ofstdyn$.
Similarly to the shift parameter, the linear-in-frequency property again yields an inharmonic reverb tail.

\subsubsection{Brick-Wall EQ}

We extend the reverb-tail algorithm with a brick-wall bandpass equalizer that can be used to tailor the frequency range of the reverb tail using the parameters \param{low-cut} and \param{high-cut}.
The equalizer is implemented in the frequency domain where it sets the noise bins outside the passband to zero
\begin{align}
\hat{u}_k &= 0,~k \in \{1,\ldots,K_\textrm{HP}, K_\textrm{LP}+1,\ldots,K\}.
\end{align}

\subsubsection{Stereo Options}

So far, we have only detailed mono audio processing.
We extend \fdverb to stereo signals by running two tail-reverberator instances in parallel, one per channel.
The \param{width} parameter controls the correlation of the noise spectrum $\hat\bmu$ between the two channels, ranging from fully correlated (mono, i.e., \emph{mid} only), to fully uncorrelated (full stereo, i.e., \emph{mid} and \emph{side}), to fully anti-correlated (out-of-phase, i.e., \emph{side} only).
The \param{link} parameter controls whether the envelopes of the two channels are processed independently or averaged into a shared mono envelope.

Furthermore, the early-reflection taps are distributed across the stereo field using amplitude panning. The panning follows a fixed random pattern, with the \param{width} parameter controlling the spread.

\section{DAW Plug-In Implementation}
\label{sec:plugin}

\fdverb is implemented as a plug-in for digital audio workstations (DAWs) using the JUCE framework\footnote{\url{https://github.com/juce-framework/JUCE}} and is provided in the VST3 and Audio Unit (AU) formats. 
The ER unit is implemented using a delay line, and up to $32$ echoes are supported with a maximum delay of \qty{500}{\ms}.
The spectral transform from \fref{sec:forward_transform} is implemented using the real fast Fourier transform (RFFT) provided by JUCE and block sizes from $N = 2^5$ to $N = 2^{14}$ are supported.
To support DAW buffer sizes larger and smaller than the block size~$N$, the tail reverberator runs on a dedicated background thread, communicating with the real-time audio callback through lock-free queues.
In the current version, we measured the average per-block reverb-tail processing time to range from \qty{0.011}{\ms} at $N=2^5$ to \qty{3.81}{\ms} at $N=2^{14}$.
Across all supported block sizes $N$, the per-track CPU meter of REAPER reported less than \qty{2.6}{\percent} for one instance of the \fdverb plug-in.\footnote{Measurements performed on a MacBook Pro (M2 Pro), \fdverb VST3 in REAPER v7.76, $512$ samples I/O buffer size, $\fs = \qty{48}{\kHz}$; the CPU meter reading is taken from REAPER's FX window.}

\begin{figure}[t]
\centering
\includegraphics[width=0.95\columnwidth]{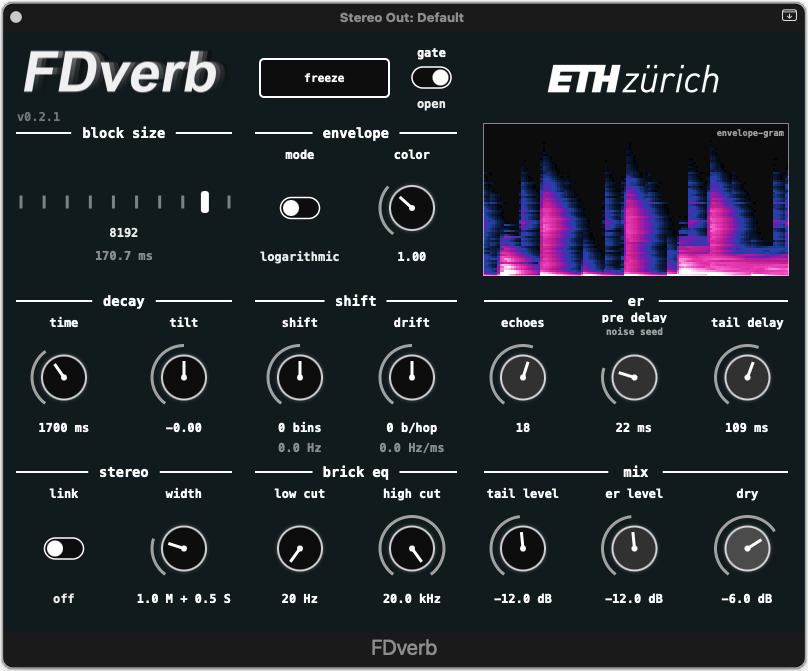}
\caption{\label{fig:plugin_ui}User interface of the FDverb plug-in with the real-time envelope visualization found at the top right.}
\end{figure}

The user interface of the \fdverb plug-in features all of the parameters mentioned in \fref{sec:algorithm} and also adds a real-time ``envelope-gram'' of the envelopes $\envbin$, a visualization similar to a rolling spectrogram; see \fref{fig:plugin_ui} for a screenshot of the user interface.
At the time of writing, the \fdverb plug-in described here is version \texttt{v1.0} of the Git repository.

\section{Visitor Experience}

Visitors are invited to explore \fdverb hands-on via a laptop running the DAW plug-in, connected to an audio interface and headphones.
We provide a set of parameter presets for \fdverb as a starting point, and visitors are encouraged to tweak the parameters in real time.
As an audio input source, visitors can choose from pre-recorded instrument samples and music, or connect their own device.




\bibliographystyle{IEEEtranDAFx}
\bibliography{26DAFx_FDverb}

\begin{thebibliography}{1}
\providecommand{\url}[1]{#1}
\csname url@samestyle\endcsname
\providecommand{\newblock}{\relax}
\providecommand{\bibinfo}[2]{#2}
\providecommand{\BIBentrySTDinterwordspacing}{\spaceskip=0pt\relax}
\providecommand{\BIBentryALTinterwordstretchfactor}{4}
\providecommand{\BIBentryALTinterwordspacing}{\spaceskip=\fontdimen2\font plus
\BIBentryALTinterwordstretchfactor\fontdimen3\font minus
  \fontdimen4\font\relax}
\providecommand{\BIBforeignlanguage}[2]{{%
\expandafter\ifx\csname l@#1\endcsname\relax
\typeout{** WARNING: IEEEtran.bst: No hyphenation pattern has been}%
\typeout{** loaded for the language `#1'. Using the pattern for}%
\typeout{** the default language instead.}%
\else
\language=\csname l@#1\endcsname
\fi
#2}}
\providecommand{\BIBdecl}{\relax}
\BIBdecl

\bibitem{moorer1979about}
J.~A. Moorer, ``About this reverberation business,'' \emph{Computer Music J.},
  vol.~3, no.~2, pp. 13\,--\,18, Jun. 1979.

\bibitem{vickers2006frequency}
E.~Vickers, J.-L.~L. Wu, P.~G. Krishnan, and R.~N.~K. Sadanandam, ``Frequency
  domain artificial reverberation using spectral magnitude decay,'' in
  \emph{Proc. 121st Audio Eng. Soc. Conv.} Audio Eng. Soc., Oct. 5\,--\,8,
  2006.

\bibitem{vilkamo2010sparse}
J.~Vilkamo, B.~Neugebauer, and J.~Plogsties, ``Sparse frequency-domain
  reverberator,'' in \emph{Proc. 40th Intl. Audio Eng. Soc. Conf. on Spatial
  Audio}. Audio Eng. Soc., Oct. 8\,--\,10, 2010.

\bibitem{waproduction_venom}
{W. A. Production}. Venom. {A}ccessed: Apr. 10, 2026. [Online]. {A}vailable:
  \url{https://www.waproduction.com/plugins/view/venom}.

\bibitem{dafx2025graetli}
J.~Roth, S.~Krebs, and C.~Studer, ``{graetli}: A microcontroller-based {DSP}
  platform for real-time audio signal processing,'' in \emph{Intl. Conf.
  Digital Audio Effects (DAFx25)}, Ancona, Italy, Sept. 2\,--\,5, 2025, (demo
  paper).

\bibitem{zolzer2011dafx}
U.~Z{\"o}lzer, Ed., \emph{{DAFx}: Digital Audio Effects}, 2nd~ed. Chichester,
  UK: John Wiley \& Sons, 2011.

\bibitem{mcnally1984dynamic}
G.~W. McNally, ``Dynamic range control of digital audio signals,'' \emph{J.
  Audio Eng. Soc.}, vol.~32, no.~5, pp. 316--327, 1984.

\end{thebibliography}

\end{document}